\documentclass[a4paper,10pt]{article}

\usepackage{a4wide}

\usepackage{amsmath}
\usepackage{amsfonts}
\usepackage{amssymb}

\usepackage{cite}

\usepackage{graphicx}

\usepackage{hyperref}

\newcommand{\bm}{\mathbf}

\title{Negative index of refraction, spacetime folding and perfect imaging in transformation optics}
\author{Luzi Bergamin\thanks{Email: Luzi.Bergamin@tkk.fi}\\ \begin{small} Department of Radio Science and Technology\end{small}\\ \begin{small}Helsinki University of Technology, FI--02015 TKK, Finland\end{small}}
\date{\today}

\begin{document}

\maketitle

\begin{abstract}
Negative index of refraction has become an accepted part of transformation optics, which is encountered in transformations that change the orientation of the manifold. Based on this concept, various designs of perfect lenses have been proposed, which all rely on a folding of space or spacetime, where the maps from electromagnetic space to laboratory space are multi-valued. Recently, a new concept for perfect imaging has been proposed by Leonhardt and Philbin, which also uses multi-valued maps, but does neither include negative index of refraction nor an amplification of evanescent modes. In this context it was speculated that multi-valued maps should be seen as the basis of perfect imaging rather than amplification of evanescent modes. It might be useful to review the standard lens based on negative index of refraction from this point of view. In this paper we show that a negative index of refraction is not an inherent characteristic of transformation optics, but rather appears as a specific choice of a sign ambiguity. Furthermore, we point out that the transformation designed lens does not amplify evanescent modes, in contrast to the Pendry-Veselago lens. Instead, evanescent modes at the image point are produced by a duplicated source and thus no imaging of the near field takes place.
\end{abstract}

\section{Introduction}
Negative index of refraction and perfect lenses \cite{Pendry:Pl2000,Veselago:1968Nr,Ramakrishna:2005Rp,Veselago:2006Nr} have become one of the most important concepts in metamaterials. The theoretical design of such devices was considerably stimulated by the observation \cite{Leonhardt:2006Nj} that a negative index of refraction can be understood from transformation optics as a transformation of space that inverts its orientation. Based on this idea, not only the flat perfect lens was re-interpreted as a folding of space \cite{Leonhardt:2006Nj}, but also lenses with different shapes \cite{Pendry:2003Jp,Pendry:2003Cl,Yan:2008Pr,Bergamin:2008Mm} were proposed. In all these concepts a perfect lens is established by folding of space, such that three points in laboratory space (one on each side of the lens and one inside the lens) correspond to a single point in the virtual electromagnetic space that is used to derive the media properties. Based on these successes it was natural to conclude that transformation optics is an ideal tool to design perfect imaging devices. Recently, it was suggested \cite{Leonhardt:2009Td,Leonhardt:2009Pi} that perfect imaging should rather be seen as the result of multi-valued maps than an effect of the amplification of evanescent waves. These results suggest to critically review the role of negative index of refraction and perfect lenses within transformation optics. In Sec.~\ref{se:negativerefr} it is reviewed how negative values of permittivity and permeability can emerge in transformation optics. It is pointed out that a negative index of refraction cannot be seen as an inherent characteristic of transformation optics similar to the bending of light as used in cloaking \cite{Pendry:2006Sc,Leonhardt:2006Sc}. Rather, these values are obtained from a clever choice of signs in an ambiguity related to orientation changing transformations. Sec.~\ref{se:interface} presents an argument that can justify the choice of conventions that yields to negative refraction. Still, as will be shown in Sec.~\ref{se:lenses}, results from transformation optics based on multi-valued maps should be used with utmost care. In particular, the transformation optics analogue of the Pendry-Veselago lens neither amplifies evanescent modes nor includes an imaging of the near field.

\section{Negative index of refraction in transformation optics}
\label{se:negativerefr}
\begin{figure}[t]
 \centering
 \includegraphics[width=0.8\linewidth]{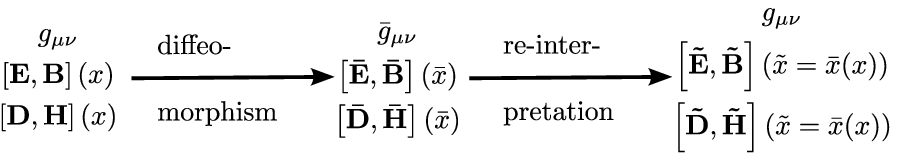}
 \caption{Illustration and notation of transformation optics.}
 \label{fig:TOnotation}
\end{figure}

It is the purpose of this section to review how a negative index of refractive appears in transformation optics. In the logic of transformation optics one starts by writing down a vacuum solution $\bm D = \varepsilon_0 \bm E$, $\bm B = \mu_0 \bm H$ of the Maxwell equations\footnote{Transformation optics relies on generic coordinates and thus an appropriate formalism has to be employed. Here, we follow Refs.\ \cite{Leonhardt:2006Nj,Leonhardt:2008Oe,Bergamin:2008Pa} and use component notation in conjunction with the Einstein summation convention. Thus in all equations a summation over repeated indices is assumed. Latin indices refer to space and the sum is performed over the values $i=1,2,3$. Greek indices are spacetime indices, the sum runs over $\mu=0,1,2,3$, whereby $x^0 = c t$ is interpreted as time. Further explanations on our notations and conventions can be found in the appendix.}
\begin{align}
\label{maxwell}
 \nabla_i B^i &= 0\ , & \nabla_0 B^i + \epsilon^{ijk}\partial_j E_k &= 0\ , \\
\label{maxwell2}
 \nabla_i D^i &= \rho\ , & \epsilon^{ijk} \partial_j H_k - \nabla_0 D^i &= j^i\ .
\end{align}
To account for possibly curvilinear coordinates we used the the covariant derivative in three dimensions, $\nabla_i$, with
\begin{equation}
\label{covder}
 \nabla_i A^i = (\partial_i + \Gamma^i_{i j}) A^j = \frac{1}{\sqrt{\gamma}} \partial_i(\sqrt{\gamma}A^i)\ ,
\end{equation}
where $\gamma$ is the determinant of the space metric $\gamma_{ij}$. Now a diffeomorphism to a virtual space called electromagnetic space is defined, which locally is implemented as a coordinate transformation $x^i \rightarrow \bar x^i(x)$. Its effect is captured by re-writing the Maxwell equations in terms of the new, barred variables. More involved is the new relation among the fields $\bar{ \bm D}$, $\bar{ \bm B}$, $\bar{ \bm E}$ and $\bar{ \bm H}$, which in a generic coordinate system takes the form \cite{Landau2}
\begin{align}
\label{epsorig}
 \bar D^i &= \varepsilon_0 \frac{\bar \gamma^{ij}}{\sqrt{-\bar g_{00}}} \bar E_j - \frac{\bar g_{0j}}{\bar g_{00} c} \bar \epsilon^{jil} \bar H_l\ , &
 \bar B^i &= \mu_0 \frac{\bar \gamma^{ij}}{\sqrt{-\bar g_{00}}} \bar H_j + \frac{\bar g_{0j}}{\bar g_{00} c} \bar \epsilon^{jil} \bar E_l\ .
\end{align}
Here, $\bar g_{\mu\nu}$ are the components of the transformed spacetime metric, from which the transformed space metric follows as $\bar \gamma^{ij} = \bar g^{ij}$. These relations resemble the constitutive relations of a special medium, but of course just describe the same physics as Eqs.~\eqref{maxwell} and \eqref{maxwell2}, re-written in complicated coordinates. To make use of the relations \eqref{epsorig} as media parameters, the solutions $\bar{ \bm D}$, $\bar{ \bm B}$, $\bar{ \bm E}$ and $\bar{ \bm H}$ are turned back into solutions in terms of the metric $g_{\mu\nu}$ in the coordinate system $x^\mu$, while keeping the form of the ``constitutive relations'' \eqref{epsorig} in terms of $\bar g_{\mu\nu}$. Since the Maxwell equations only depend on the determinant of the metric, but not on its specific components, this can be achieved by the simple rescaling \cite{Leonhardt:2006Nj,Bergamin:2008Pa}
\begin{align}
\label{scal1}
 \tilde{\bm E} &= \bar{\bm E}\ , & \tilde{\bm B} &= \frac{\sqrt{\bar \gamma}}{\sqrt{\gamma}} \bar{\bm B}\ , &
 \tilde{\bm D} &= \frac{\sqrt{\bar \gamma}}{\sqrt{\gamma}}  \bar{\bm D}\ , & \tilde{\bm H} &= \bar{\bm H}\ .
\end{align}
If  $\bar{ \bm D}$, $\bar{ \bm B}$, $\bar{ \bm E}$ and $\bar{ \bm H}$ are a solution of the Maxwell equations with metric $\bar g_{\mu\nu}$ and with ``constitutive relations'' \eqref{epsorig}, then $\tilde{ \bm D}$, $\tilde{ \bm B}$, $\tilde{ \bm E}$ and $\tilde{ \bm H}$ are a solution in terms of the coordinates $x^\mu$ with metric $g_{\mu\nu}$ and with a constitutive relation
\begin{align}
\label{epstilde2}
 \tilde{ D}^i &= \varepsilon_0 \frac{\bar g^{ij}}{\sqrt{-\bar g_{00}}}  \frac{\sqrt{\bar \gamma}}{\sqrt{\gamma}} \tilde E_j - \frac{\bar g_{0j}}{\bar g_{00}c} \epsilon^{jil} \tilde{ H}_l\ , &
 \tilde B^i &= \mu_0 \frac{\bar g^{ij}}{\sqrt{-\bar g_{00}}} \frac{\sqrt{\bar \gamma}}{\sqrt{\gamma}} \tilde{ H}_j + \frac{\bar g_{0j}}{\bar g_{00}c} \epsilon^{jil} \tilde E_l\ .
\end{align}
In contrast to Eq.\ \eqref{epsorig}, which still describe electrodynamics in empty space, the constitutive relations \eqref{epstilde2} describe electrodynamics in a medium. The basic idea of transformation optics is illustrated in Fig.\ \ref{fig:TOnotation}, which also summarizes our notation.

An additional problem within the program of transformation optics has been pointed out by Leonhardt and Philbin \cite{Leonhardt:2006Nj}: Consider a transformation of the coordinates that changes the orientation of the manifold, i.e.\ that maps a right-handed coordinate system onto a left-handed one and vice versa. Examples of such transformations are $\bar z = -z$ or the combined transformation $\bar x = y$, $\bar y = x$. Let's consider the second example and assume that we start with a right-handed coordinate system, $\epsilon^{xyz} = 1/\sqrt{\gamma}$. Since the transformation does not change the determinant of the metric, $\gamma = \bar \gamma$, we immediately obtain
\begin{equation}
 \bar \epsilon^{\bar y\bar x\bar z} = \epsilon^{\tilde y \tilde x \tilde z} = 1/\sqrt{\gamma}\ ,
\end{equation}
where in the first step the coordinate transformation has been applied and in the second step the result has been re-interpreted in terms of the original space. Obviously, under transformations that change the orientation of the manifold the sign of the anti-symmetric tensor changes as well. Thus, they lead to a sign error in the Maxwell equations since the cross products change sign and we have to conclude that the recipe \eqref{scal1} does not yield solutions of the correct Maxwell equations if such orientation changing transformations are allowed.

To circumvent this problem it has been suggested in Ref.\ \cite{Leonhardt:2006Nj} to include an additional sign in the rescalings \eqref{scal1}. Indeed, with the new definition
\begin{align}
\label{scal3}
 \tilde{\bm E} &= \bar \sigma \bar{\bm E}\ , & \tilde{\bm B} &= \frac{\sqrt{\bar \gamma}}{\sqrt{\gamma}} \bar{\bm B}\ , &
 \tilde{\bm D} &= \frac{\sqrt{\bar \gamma}}{\sqrt{\gamma}} \bar{\bm D}\ , & \tilde{\bm H} &= \bar \sigma \bar{\bm H}\ ,
\end{align}
with
\begin{equation}
 \bar \sigma = \begin{cases}
           +1 & \text{if $x^i \rightarrow \bar x^i$ does not change the orientation,}\\
	   -1 & \text{otherwise,}
          \end{cases}
\end{equation}
these signs are absorbed in the definition of $\tilde{\bm E}$ and $\tilde{\bm H}$. But these new re-interpretations of the solutions also affect the constitutive relations \eqref{epstilde2}, which now become
\begin{align}
\label{epstilde3}
 \tilde{ D}^i &= \bar \sigma\frac{\bar g^{ij}}{\sqrt{-\bar g_{00}}}  \frac{\sqrt{\bar \gamma}}{\sqrt{\gamma}} \tilde E_j - \frac{\bar g_{0j}}{\bar g_{00}} \epsilon^{jil} \tilde{ H}_l\ , &
 \tilde B^i &= \bar \sigma\frac{\bar g^{ij}}{\sqrt{-\bar g_{00}}} \frac{\sqrt{\bar \gamma}}{\sqrt{\gamma}} \tilde{ H}_j + \frac{\bar g_{0j}}{\bar g_{00}} \epsilon^{jil} \tilde E_l\ .
\end{align}
This leads to the important conclusion that \emph{transformations that change the orientation of the coordinate system yield media with negative index of refraction.}

Since this conclusion is intimately connected with the spacetime symmetries of electrodynamics, it makes sense to review the argument in terms of a spacetime covariant formulation of electrodynamics. Thus, the fields $\bm E$ and $\bm B$ are combined to the field strength tensor $F_{\mu\nu}$, while $\bm D$ and $\bm H$ become parts of the excitation tensor $\mathcal H^{\mu\nu}$. The space vectors are found by the identifications
\begin{align}
\label{Fident}
 E_i &= F_{0i}\ , & B^i &= -\frac{1}{2c} \epsilon^{ijk} F_{jk}\ ,\\
 \label{Hident}
 D^i &= - \epsilon_0 \sqrt{-g_{00}} \mathcal H^{0i}\ , & H_i &= - \frac{\epsilon_0c \sqrt{-g_{00}}}{2} \epsilon_{ijk} \mathcal H^{jk}
\end{align}
and the Maxwell equations are rewritten as
\begin{align}
\label{faraday}
 \epsilon^{\mu\nu\rho\sigma}\partial_{\nu} F_{\rho\sigma}&=0\ , & & \text{Maxwell-Faraday equations;}\\
 \label{ampere}
 D_{\mu} \mathcal{H}^{\mu\nu} = \frac{1}{\sqrt{-g}} \partial_\mu (\sqrt{-g} \mathcal H^{\mu\nu})&= J^\nu\ , && \text{Maxwell-Amp\`ere equations.}
\end{align}
Here, $D_{\mu}$ is the covariant spacetime derivative \eqref{covderII} and $J^\mu$ is the four-current encompassing $\rho$ and $\bm j$. 
This formulation has the advantage that all spacetime symmetries are manifest and consequently spatial, time and mixed spacetime transformations can be treated on the same footing. While the Maxwell-Amp\`ere equations are invariant under any change of orientation, the Maxwell-Faraday equations change sign if \emph{spacetime} changes its orientation due to the four-dimensional anti-symmetric tensor $\epsilon^{\mu\nu\rho\sigma}$. Thus, it has been suggested in Ref.~\cite{Bergamin:2008Pa} that the field strength tensor in the medium should be defined as
\begin{equation}
 \tilde F_{\mu\nu} = \bar s \bar F_{\mu\nu}\ ,
\end{equation}
where $\bar s$ now indicates a change in the orientation of \emph{spacetime} rather than just space. Thus, also a map $\bar t = -t$ yields a negative index of refraction. Consequently, the signs $\bar \sigma$ in Eq.\ \eqref{epstilde3} have to be replaced by $\bar s$ in this prescription.

On second thought, however, it is seen that an eventual change of sign in the Maxwell-Faraday equations remains without consequences, simply since the sign just appears as an overall factor. Thus, even for orientation changing transformations \emph{the correct Maxwell equations in the medium are found without sign ambiguity and there is no need for a negative index of refraction.}

The apparent contradiction is resolved immediately by looking at Eqs\ \eqref{Fident} and \eqref{Hident}. In these equations the magnetic fields $\bm B$ and $\bm H$ are unambiguously defined as pseudo-vectors and thus they are reversed if the orientation of the manifold changes. Thus, although no signs appear in the maps of spacetime tensors, they reappear in the maps of space vectors:
\begin{align}
\label{scal5}
 \tilde{\bm E} &=  \bar{\bm E}\ , & \tilde{\bm B} &= \bar \sigma \frac{\sqrt{\bar \gamma}}{\sqrt{\gamma}} \bar{\bm B}\ , &
 \tilde{\bm D} &= \frac{\sqrt{\bar \gamma}}{\sqrt{\gamma}} \bar{\bm D}\ , & \tilde{\bm H} &= \bar \sigma  \bar{\bm H}\ .
\end{align}
Since this prescription changes the signs of $\bm B$ and $\bm H$ in case of orientation changing transformations, the constitutive relations \eqref{epstilde2} are not affected and consequently there is no room for negative index of refraction.

This exercise shows that some sign changes are unavoidable in the re-interpretation of the transformed vacuum solutions as medium solutions due to the cross product in the Maxwell equations and the fact that $\bm E$ and $\bm D$ are vectors, while $\bm B$ and $\bm H$ are pseudo-vectors. However, there are different ways treat these signs and depending on the choice different maps yield negative index of refraction.

The three prescriptions presented here are summarized in Table \ref{tab:summary}, it should be noted that this list is not exhaustive, but further possibilities to distribute the signs are conceivable.
\begin{table}[t]
\begin{center}
\[
 \begin{array}{|l|ccc|}
  \hline&&&\\[-1.6ex]
  \mbox{Type} & \mbox{(A): \cite{Leonhardt:2006Nj,Leonhardt:2008Oe}} & \mbox{(B): \cite{Bergamin:2008Pa}} & \mbox{(C): minimal}\\ \hline&&&\\[-1.6ex]
  \mbox{Spacetime Tensors} & \tilde F_{\mu\nu} = \bar\sigma F_{\mu\nu} & \tilde F_{\mu\nu} = \bar s F_{\mu\nu} & \tilde F_{\mu\nu} = F_{\mu\nu} \\
   & \tilde{\mathcal H}^{\mu\nu} = \frac{\sqrt{-\bar g}}{\sqrt{-g}} \bar{\mathcal H}^{\mu\nu} & \tilde{\mathcal H}^{\mu\nu} = \frac{\sqrt{-\bar g}}{\sqrt{-g}} \bar{\mathcal H}^{\mu\nu} & \tilde{\mathcal H}^{\mu\nu} = \frac{\sqrt{-\bar g}}{\sqrt{-g}} \bar{\mathcal H}^{\mu\nu} \\ \hline&&&\\[-1.6ex]
  \mbox{Space Vectors} & \tilde E_i =  \bar \sigma \bar E_i & \tilde E_i =  \bar s \bar E_i & \tilde E_i =  \bar E_i \\
  & \tilde D^i =  \frac{\sqrt{\bar \gamma}}{\sqrt{\gamma}} \bar D^i & \tilde D^i =  \frac{\sqrt{\bar \gamma}}{\sqrt{\gamma}} \bar D^i & \tilde D^i =  \frac{\sqrt{\bar \gamma}}{\sqrt{\gamma}} \bar D^i \\\hline&&&\\[-1.6ex]
  \mbox{Space Pseudo-Vectors} & \tilde B^i = \frac{\sqrt{\bar \gamma}}{\sqrt{\gamma}} \bar B^i & \tilde B^i = \bar s \bar \sigma\frac{\sqrt{\bar \gamma}}{\sqrt{\gamma}} \bar B^i & \tilde B^i = \bar \sigma \frac{\sqrt{\bar \gamma}}{\sqrt{\gamma}} \bar B^i \\
  & H_i = \bar \sigma \bar H_i & H_i = \bar \sigma \bar H_i & H_i = \bar \sigma \bar H_i \\ \hline&&&\\[-1.6ex]
  \mbox{Negative Index} & x^i\rightarrow \bar x^i &  x^\mu\rightarrow \bar x^\mu & \mbox{never} \\
  \mbox{of Refraction} & \mbox{changes orientation} & \mbox{changes orientation} & \\ \hline
 \end{array}
\]
%
  \end{center}
 \caption{Summary of the three discussed options to treat orientation changing transformations. $\bar \sigma$ takes value $-1$ if the spatial coordinate system changes orientation, $\bar s = -1$ if the spacetime coordinates change orientation.}
 \label{tab:summary}
\end{table}
\section{Boundary conditions and reflectionless interfaces}
\label{se:interface}
Which of the three options (A)--(C) is the correct one? There exists no definite answer to this question, there even exist more possibilities than presented here. Since the re-interpretation of the solutions $\bar F_{\mu\nu}\rightarrow \tilde F_{\mu\nu}$ and $\bar{\mathcal{H}}^{\mu\nu}\rightarrow\tilde{\mathcal{H}}^{\mu\nu}$ are an \emph{ad-hoc} manipulation, there exist no strict rules or even mathematical definitions how this should be done. As only requirement, the fields with a tilde must constitute a solution of the Maxwell equations in the space with coordinates $x^{\mu}$ and a trivial transformation $\bar x^\mu \equiv x^\mu$ must map the original solution onto itself.

From a purely mathematical point of view, option (C) clearly appears as the preferable one, since it contains no sign ambiguity at all in the spacetime formulation, or---in terms of space vectors---changes the signs of $\bm B$ and $\bm H$, which are pseudo-vectors and thus have to be odd under a change of orientation of space. Still, the possibility to describe materials with negative index of refraction is an important asset of options (A) and (B). At this point, one should remember that negative refractive index media are interesting only if they include some interfaces to normal media or empty space. Thus, it has to be studied how the different prescriptions match the boundary conditions that have to hold at such an interface.

Here, we answer this question from the point of view taken in Ref.\ \cite{Bergamin:2009In}: on both sides of an interface between two transformation media, the solution of the Maxwell equations in the medium (i.e.\ the solutions with a tilde) can be described by means of vacuum solutions (the solutions without tilde or bar.) We thus can ask the question, under which restrictions of the transformations the boundary conditions at the interface are met if the \emph{same} vacuum solution $\bm D = \varepsilon_0 \bm E$, $\bm B = \mu_0 \bm H$ is used on both sides. This provides a sufficient condition for a reflectionless interface and in addition guarantees that the interpretation of transformation optics as ``mimicking a different space'' indeed extends across the interface.

The transformations as presented in Table \ref{tab:summary} are not yet sufficient to perform this task, but we need the expressions of the medium solutions in terms of the original vacuum solution. In the following we exclude bi-anisotropic media; then the transformations for $\bm D$ and $\bm H$ are \cite{Bergamin:2009In}
\begin{align}
 \label{tildeDH2}
 \tilde D^i\left(\tilde x = \bar x(x)\right) &= \alpha \bar s \left\lvert \frac{\partial x^k}{\partial \bar x^l}\right\lvert \frac{\partial \bar x^i}{\partial x^j} D^j(x) \ , &
 \tilde{H}_i\left(\tilde x = \bar x(x)\right) &= \alpha \bar s \frac{\partial x^0}{\partial \bar x^0} \frac{\partial x^j}{\partial \bar x^i} H_j(x)\ ,
\end{align}
while the transformations of $\bm E$ and $\bm B$ depend on the chosen prescription (A)--(C):
\begin{align}
\label{tildeEBA}
 (A): && \tilde{E}_i\left(\tilde x = \bar x(x)\right) &= \alpha \bar \sigma \frac{\partial x^0}{\partial \bar x^0} \frac{\partial x^j}{\partial \bar x^i} E_j(x)\ , & \tilde B^i\left(\tilde x = \bar x(x)\right) &= \alpha \bar \sigma \left\lvert \frac{\partial x^k}{\partial \bar x^l}\right\lvert \frac{\partial \bar x^i}{\partial x^j} B^j(x) \ , \\
 \label{tildeEBB}
 (B): && \tilde{E}_i\left(\tilde x = \bar x(x)\right) &= \alpha \bar s \frac{\partial x^0}{\partial \bar x^0} \frac{\partial x^j}{\partial \bar x^i} E_j(x)\ , & \tilde B^i\left(\tilde x = \bar x(x)\right) &= \alpha \bar s \left\lvert \frac{\partial x^k}{\partial \bar x^l}\right\lvert \frac{\partial \bar x^i}{\partial x^j} B^j(x) \ , \\
 \label{tildeEBC}
 (C): && \tilde{E}_i\left(\tilde x = \bar x(x)\right) &= \alpha \frac{\partial x^0}{\partial \bar x^0} \frac{\partial x^j}{\partial \bar x^i} E_j(x)\ , & \tilde B^i\left(\tilde x = \bar x(x)\right) &= \alpha \left\lvert \frac{\partial x^k}{\partial \bar x^l}\right\lvert \frac{\partial \bar x^i}{\partial x^j} B^j(x) \ .
\end{align}
In these equations we have introduced a new parameter $\alpha$, which just represents the fact that shifting \emph{all} fields by a constant does not change the constitutive relation. Without loss of generality one can assume $\alpha = \pm 1$ and moreover $\alpha \equiv 1$ in case of a trivial map $\bar x^\mu \equiv x^{\mu}$ \cite{Bergamin:2009In}. Furthermore,  $\left\lvert \partial x^k/\partial \bar x^l\right\lvert$ is the determinant of the transformation matrix.

Let us now consider a passive interface between a ``left medium'' (index $L$) and a ``right medium'' (index $R$) with boundary conditions
\begin{align}
\label{BC1}
 (\bm D_L - \bm D_R )\cdot \bm n &= 0\ , & (\bm B_L - \bm B_R )\cdot \bm n &= 0\ , \\
 \label{BC2}
 (\bm E_L - \bm E_R) \times \bm n &= 0\ , & (\bm H_L - \bm H_R) \times \bm n &= 0\ ,
\end{align}
where $\bm n$ is a vector normal to the interface. Without loss of generality we can assume an adapted coordinate system in laboratory space, such that the direction normal to the interface is labeled by the coordinate $\tilde{\bm x}_\perp = (0,0,\tilde x^\perp)$, while the directions parallel to the interface have coordinates $\tilde{\bm x}_\parallel = (\tilde x^A,0)$, whereby the index $A$ takes values $1,2$.

As is immediately seen the four conditions \eqref{BC1} and \eqref{BC2} reduce to two restrictions on the transformation if option (B) is chosen. Then, the vacuum solution extends across the interface if
\begin{align}
\label{parallelI}
 \alpha_L \bar s_L \frac{\partial x^0}{\partial \bar x_L^0} \frac{\partial x^j}{\partial \bar x_L^A} &= \alpha_R \bar s_R \frac{\partial x^0}{\partial \bar x_R^0} \frac{\partial x^j}{\partial \bar x_R^A}\ , \\
 \label{normalI}
 \alpha_L \bar s_L \left\lvert \frac{\partial x^k}{\partial \bar x_L^l}\right\lvert \frac{\partial \bar x_L^\perp}{\partial x^j} &= \alpha_R \bar s_R \left\lvert \frac{\partial x^k}{\partial \bar x_R^l}\right\lvert \frac{\partial \bar x_R^\perp}{\partial x^j}\ .
\end{align}
These restrictions are satisfied if the two transformations obey \cite{Bergamin:2009In}
\begin{align}
\label{simple}
 \frac{\partial \bar x^A_L}{\partial \bar x^B_R} &= \delta^A_B\ , & \frac{\partial \bar x^0_L}{\partial \bar x^0_R} &= 1\ , & \alpha_L &= \bar s_L\ , & \alpha_R &= \bar s_R \ .
\end{align}
$\partial \bar x_L^\perp/\partial \bar x_R^\perp$ remains unrestricted, in particular $\partial \bar x_L^\perp/\partial \bar x_R^\perp < 0$ is permitted and yields negative index of refraction. These conditions say that the transformed coordinates parallel to the interface must agree on both sides, while the transformation in the orthogonal direction is continuous, but not necessarily differentiable at the interface. Furthermore, the time coordinates must agree on both sides. A negative index of refraction results as an inversion of the direction normal to the interface.

Contrariwise, options (A) and (C) yield four different conditions. Still, for all cases that meet the restrictions \eqref{simple} option (A) reduces to the case (B), since time inversions are excluded. Thus $\bar \sigma \equiv \bar s$ in all cases that allow an extension of the vacuum solution across the interface. Still, within the prescription (C) the boundary conditions \eqref{BC1} and \eqref{BC2} cannot be met with the same vacuum solution on both sides of the interface unless $\bar s_L \equiv \bar s_R$, i.e.\ there is no change of orientation.

We thus conclude that there exist good reasons to chose options (A) or (B) since these prescriptions allow to describe a larger class of interfaces by means of a single vacuum solution than option (C) and thereby also allow to describe negative refractive index materials.

\section{Perfect lenses and evanescent modes}
\label{se:lenses}
If negative index of refraction can be made part of transformation optics, how good is this interpretation? To our knowledge perfect lenses \cite{Pendry:Pl2000} are the only device where transformation optics with negative index of refraction was proposed \cite{Leonhardt:2006Nj}. Thus, we restrict to this example here. A flat lens is associated with the map \cite{Leonhardt:2006Nj}
\begin{equation}
\label{lensmap}
  z = \begin{cases}
           \bar z\ , &  \bar z<0\ ; \\
	   -\alpha \bar z\ , & 0< \bar z<D\ ; \\
	   \bar z-(\alpha+1) D\ , &  \bar z>D\ .
          \end{cases}
\end{equation}
As is easily seen, any point $-\alpha D< z < 0$ is mapped on three different points in the virtual electromagnetic space and---upon re-interpretation---on three different points in laboratory space, whereby in the region $0<\tilde z<\alpha D$ a medium with negative index of refraction emerges. This triple valued map was associated with perfect imaging, since any solution of the Maxwell equations in the region $-\alpha D< -z_S < 0$ is reproduced exactly inside the lens at $\tilde z = z_S/\alpha$ and on the other side of the lens at $\tilde z = (\alpha+1)D - z_S$.
\begin{figure}[t]
 \centering
 \includegraphics[width=\linewidth,bb=0 0 965 348]{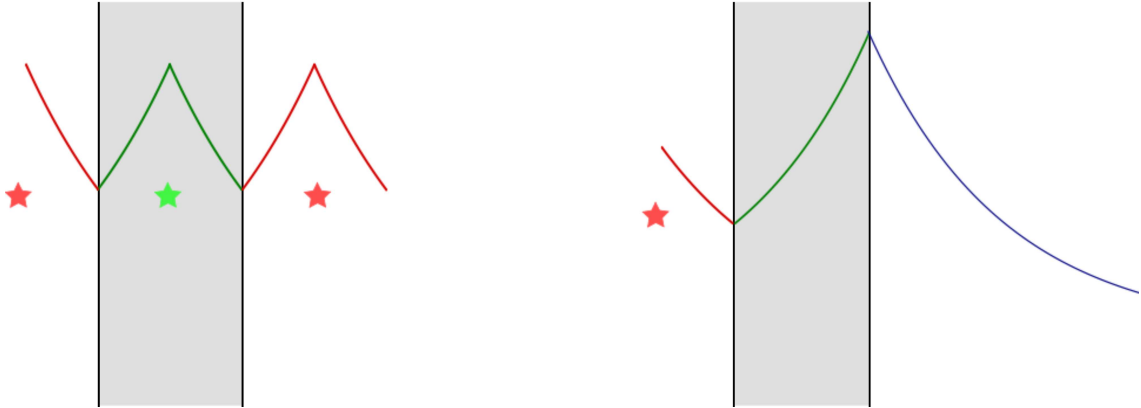}
 \caption{Schematic view of evanescent modes in the the transformation optics lens (left hand side) and the Pendry-Veselago lens (right hand side). While the latter images by amplification of evanescent modes \cite{Pendry:Pl2000,Ramakrishna:2005Rp}, the former triples the sources.}
 \label{fig:lens comparison}
\end{figure}

Since negative index of refraction within transformation optics is rather an effect of the choice of signs than an inherent characteristic, one should have a careful look at the lens proposed by the map \eqref{lensmap}. The following three conclusions are immediate:
\begin{enumerate}
 \item Due to causality, the transformation optics lens is strictly limited to stationary situations. It is well known that transformation designed concepts can get in conflict with causality, but mostly this can be resolved by a limitation to a rather narrow bandwidth. However, the folding of space by means of the map \eqref{lensmap} limits the application of this concept to strictly stationary situations, simply because any change in the electromagnetic fields at the source point causes an \emph{instantaneous} change of the mirror image inside the lens and the image behind the lens.
 \item The transformation designed lens cannot image a source, but rather triples it. Indeed, a situation with a source at the source point, but empty mirror image and image point, is not covered by transformation optics. Instead, a source automatically creates a mirror source (sink) inside the lens and a second source behind the lens (see Fig.\ \ref{fig:lens comparison}.)
 \item Consequently, within transformation optics no enhancement of the evanescent waves takes place, which is the working principle of the Pendry-Veselago lens \cite{Pendry:Pl2000,Ramakrishna:2005Rp}. As can be seen from Fig.\ \ref{fig:lens comparison}, all evanescent waves in the transformation designed lens are easily explained as the evanescent modes generated by one of the three sources. There is no need for an amplification of such modes.
\end{enumerate}

\section{Conclusions}
\label{se:conclusion}
In this paper we have reviewed the role of negative index of refraction within transformation optics. It was shown that negative refraction emerges as a consequence of a sign ambiguity and thus should not be seen as an inherent characteristic of transformation optics. Indeed, variants of transformation optics without negative refraction are consistent and follow immediately from the expected behavior of the electromagnetic fields under orientation changing transformations. Nonetheless, it can be argued that negative refraction should be included, since this formulation allows a simpler description of interfaces between different transformation media.

The most important application of a negative index of refraction, the perfect lens, has shortly been reviewed starting from the above observation. Most importantly it was found that the transformation designed lens does not amplify the evanescent modes, but at the same time also is unable to image a source. This observation might be important with respect to recent ideas on perfect imaging without negative refraction \cite{Leonhardt:2009Pi,Leonhardt:2009Td}. Also this concept relies on multi-valued maps, notice however that in these works several points in electromagnetic space are mapped onto a single point in laboratory space, rather than vice versa as in the case of the lens discussed here. Thus, even within transformation optics this concept is not restricted to stationary situations.

As a general conclusion it should be stressed that transformation designed imaging devices should be used with utmost care, in particular if they include negative index of refraction. In many cases, the analysis essentially is restricted to stationary situations where sources are not imaged, but rather duplicated. Consequently, such concepts do not amplify evanescent modes, but they are produced at source and image points by means of the multiple sources.

\subsection*{Acknowledgment}
The author would like to thank S.~Tretyakov, C.~Simovski, I.~Nefedov, P.~Alitalo and A.~Favaro for stimulating discussions. This project was supported by the Academy of Finland, project no.\ 124204.

\appendix
\section{Covariant formulation}
In this Appendix we present our notations and conventions regarding the covariant formulation the Maxwell equations on a generic (not necessarily flat) manifold and written in general coordinates. For a detailed introduction to the topic we refer to the relevant literature, e.g.\ \cite{Landau2,Post}.

Greek indices $\mu, \nu, \rho, \ldots$ are spacetime indices and run from 0 to 3, Latin indices $i,j,k,\ldots$ space indices with values from 1 to 3. Furthermore an adapted coordinate system is used at the interface, such that $(x^i) = (x^A,x^\perp)$, where $x^A$ are the directions parallel to the interface, while $x^\perp$ is perpendicular. Therefore capital Latin indices take values 1,2.
 
For the metric we use the ``mostly plus'' convention, so the standard flat metric is $g_{\mu\nu} = \mbox{diag}(-1,1,1,1)$. Time is always interpreted as the zero-component of $x^{\mu}$, $x^0 = c t$. With this identification an induced space metric can be obtained as \cite{Landau2}
\begin{equation}
\label{indmetric}
 \gamma^{ij} = g^{ij}\ , \qquad \gamma_{ij} = g_{lk} - \frac{g_{0i}g_{0j}}{g_{00}}\ , \qquad \gamma^{ij} \gamma_{jk} = \delta^i_k\ ,
\end{equation}
where $\delta^i_k$ is the Kronecker symbol.
This implies as relation between the determinant of the spacetime metric, $g$, and the one of the space metric, $\gamma$,
\begin{equation}
\label{detrel}
 -g = -g_{00} \gamma\ .
\end{equation}

In the relativistically covariant formulation $\bm E$ and $\bm B$ are combined to the field strength tensor $F_{\mu\nu}$, while $\bm D$ and $\bm H$ become part of the excitation tensor $\mathcal H^{\mu\nu}$:
\begin{align}
 [F_{\mu\nu}]&= \begin{pmatrix}
                  0 & E_1 & E_2 & E_3\\-E_1 & 0 & -c B^3 & c B^2 \\
		  -E_2 & c B^3 & 0 & -c B^1 \\
		  -E_3 & -c B^2 & c B^1 & 0
		  \end{pmatrix} & [\mathcal H^{\mu\nu}]&= \frac{1}{\varepsilon_0 \sqrt{g_{00}}}\begin{pmatrix}
                  0 & -D^1 & -D^2 & -D^3\\ D^1 & 0 & - \frac{H_3}{c} & \frac{H_2}{c} \\
		  D^2 & \frac{H^3}{c} & 0 & -\frac{H^1}{c} \\
		  D^3 & -\frac{H_2}{c} & \frac{H_1}{c} & 0
		  \end{pmatrix}
\end{align}
Finally, electric charge and current are combined into a four-current $J^{\mu} =(\sqrt{g_00} \varepsilon_0)^{-1} (\rho, j^i/c)$. In this way the Maxwell equations can be written in the compact form
\begin{align}
\label{EOMcomp}
 \epsilon^{\mu\nu\rho\sigma} \partial_\nu F_{\rho \sigma} &= 0\ , & D_\nu \mathcal H^{\mu \nu} &= - J^\mu\ .
\end{align}
The Maxwell equations depend on the metric through the covariant derivative $D_\mu$. Since
\begin{equation}
\label{covderII}
 D_\nu \mathcal H^{\mu \nu} = (\partial_\mu + \Gamma^{\nu}_{\nu \rho}) \mathcal H^{\mu \rho} = \frac{1}{\sqrt{-g}} \partial_{\nu} (\sqrt{-g} \mathcal H^{\mu\nu})
\end{equation}
it is seen that the Maxwell equations just depend on the determinant of the metric, but not on its individual components.

Diffeomorphisms can change the orientation of a manifold, such that a right-handed coordinate system in laboratory space is mapped onto a left-handed one in electromagnetic space. This induces several changes of signs due to the Levi-Civita tensor that appears in the Maxwell equations. The four dimensional Levi-Civita tensor is defined as
\begin{align}
 \epsilon_{\mu\nu\rho\sigma} &= \sqrt{-g}[\mu\nu\rho\sigma]\ , & \epsilon^{\mu\nu\rho\sigma} &= - \frac1{\sqrt{-g}}[\mu\nu\rho\sigma]\ ,
\end{align}
with $[0123] = 1$. The relation between the four-dimensional Levi-Civita tensors two different spacetimes can be written as
\begin{equation}
 \epsilon_{\mu\nu\rho\sigma} = \bar s \frac{\sqrt{-g}}{\sqrt{-\bar g}} \bar \epsilon_{\mu\nu\rho\sigma}\ ,
\end{equation}
where $\bar s=+1$ if the corresponding map does not change the orientation of the manifold, $-1$ otherwise.
The reduction of the four dimensional to the three dimensional tensor reads
\begin{equation}
 \epsilon_{0ijk} = \sqrt{-g_{00}} \epsilon_{ijk}\ , \qquad \epsilon^{0ijk} = - \frac1{\sqrt{-g_{00}}} \epsilon^{ijk}\ .
\end{equation}
An additional complication arises in the definition of $\bar \epsilon_{ijk}$, since the orientation of the spacetime manifold may change without changing the orientation of space (e.g.\ a map $\bar t = -t$, $\bar x^i = x^i$ changes the orientation of spacetime but not of space.) Therefore the corresponding relation should be written as
\begin{align}
 \bar \epsilon_{0ijk} &= \bar s \bar \sigma \sqrt{-\bar g_{00}} \bar \epsilon_{ijk}\ , 
\end{align}
where $\bar \sigma = +1$ if the spatial part of the transformation preserves the orientation and $\bar \sigma = -1$ otherwise.

\providecommand{\href}[2]{#2}\begingroup\raggedright\endgroup
\end{document}